\documentclass[amsmath,amssymb,notitlepage]{revtex4-1}
\usepackage{dcolumn}% Align table columns on decimal point
\usepackage{bm}% bold math
\usepackage[english]{babel}
\usepackage{graphicx}
\usepackage{graphics}
\usepackage{slashed}
\graphicspath{{FIGURE/}}
\newcommand{\bb}[1]{\boldsymbol{#1}}

\begin{document}

\title{The soliton-soliton interaction in the Chiral Dilaton Model}

\author{Valentina Mantovani-Sarti$^{1}$, Byung-Yoon Park$^{2}$, Vicente Vento$^{3}$ }

\affiliation{$(1)$~Department of Physics, University of Ferrara and INFN Ferrara, $(2)$~Department of Physics, Chungam 
National University, Korea,\\ $(3)$~Department of Theoretical Physics and IFIC, University of Valencia.
}

\begin{abstract}
We study the interaction between two $B=1$ states in the Chiral Dilaton Model 
where baryons are described as non-topological solitons arising from the interaction of chiral mesons and quarks. 
By using the hedgehog solution for the $B=1$ states we construct, via a \textit{product ansatz}, three possible $B=2$ configurations 
to analyse the role of the relative orientation of the hedgehog quills in the dynamics of the soliton-soliton interaction and
investigate the behaviour of these solutions in the range of long/intermediate distance.
One of the solutions is quite binding due to the dynamics of the $\pi$ and $\sigma$ fields at intermediate distance and should be used for nuclear matter studies.
Since the product ansatz breaks down as the two solitons get close, we explore the short range distance regime with a model that describes the interaction via a six-quark bag ansatz. We calculate the interaction energy as a function of the inter-soliton distance  and
show that for small separations the six quarks bag, assuming a hedgehog structure, provides a stable bound state that at large separations connects with a special configuration coming from the product ansatz.
\end{abstract}

\maketitle

\section{Introduction}
The study of properties of nuclear matter under extreme conditions, at high temperature in relativistic heavy-ion physics and/or at high density as in compact stars, is one of the most challenging issues in hadronic physics.
At the moment the phase diagram of hadronic matter is far away from being completely defined; in fact the regime of moderate temperatures and high densities, where the non-perturbative aspects of QCD dominate, seems to present a rich set of scenarios and phases~\cite{L.McLerran2007}. 
Because of the well-known ``sign problem" in Lattice QCD \cite{Karsch:2007zz}  one cannot explore the whole phase space by the direct calculations  from the fundamental theory with  quark and gluon degrees of freedom.  Only in the last few years, it has become possible to
simulate QCD with small baryon density \cite{Fromm:2008ab}.
Chiral symmetry is a flavour symmetry of QCD which plays an essential
role in hadronic physics. At low temperatures and densities it is spontaneously
broken leading to the existence of the pion. Lattice studies seem to imply that chiral
symmetry is restored in the high temperature and/or high baryon density phases and
that it may go hand-in-hand with the confinement/deconfinement transition.
The quark condensate $\langle \bar{q} q\rangle$ of QCD is an order parameter
of this symmetry that decreases to zero when the symmetry is
restored. Thus below the phase transition, we must rely on effective field theories  defined in terms of 
hadronic fields~\cite{Weinberg:1978kz} which include the degrees of freedom and the symmetries 
relevant to the energy scale under study to guide the phenomenology of the 
QCD phase transition. These theories, so close to the QCD symmetries 
and their realizations, might be a useful laboratory to explore new phenomena close to the 
phase transition \cite{McLerran:2008ux}.

Among these theories, the Skyrme model, an effective low energy theory rooted in large $N_c$ QCD, has been applied to the dense matter studies
\cite{Park:2002ie,Lee:2003aq,Lee:2003eg,Park:2003sd,Park:2008zg,Park:2008xm,Park:2009bb}. The model does not have explicit quark and gluon degrees of freedom, and therefore one can not investigate the confinement/deconfinement transition directly, but one may study
the chiral symmetry restoration transition which occurs close by.  The main ingredient connected 
with chiral symmetry is the pion, the Goldstone boson associated with the spontaneously broken 
phase. The various patterns in which the symmetry is realized in QCD will be directly reflected in 
the in-medium properties of the pion and consequently in the properties of the skyrmions made 
of it. The classical nature of skyrmions enables one to construct the dense system quite conveniently 
by putting more and more skyrmions into a given volume. Then, skyrmions shape and arrange themselves to minimize the energy of the system. The ground state configuration of skyrmion matter are crystals. 
At low density it is made of well-localized single skyrmions \cite{Klebanov85,Goldhaber87,Castillejo89}. 
At a critical density, the system undergoes a structural phase transition to a new
kind of crystal. It is made of `half-skyrmions' which are still
well-localized but carry only  half winding number \cite{Castillejo89,Kugler89}. To implement 
the adequate realization of chiral symmetry an additional degree of freedom, associated
to the scale anomaly of QCD,  has to be incorporated, the dilaton field $\chi$ \cite{Migdal82,Ellis85}. 
The dilaton field takes over the role of the order parameter for  chiral symmetry
restoration \cite{Lee:2003eg,Park:2008zg}.

Recently, the  Chiral-Dilaton Model (CDM), with hadronic degrees of freedom, has been shown to provide a 
good description of nuclear physics at densities around $\rho_0$ and  to  describe the gradual restoration of chiral
symmetry at higher densities~\cite{Bonanno:2008tt}.  In~\cite{Drago:2011hq} the authors used this model interpreting 
the fermions not as nucleons, but as quarks and applying the \textit{Wigner-Seitz approach}~\cite{Wigner:1933zz}, they
showed that the new potential allows to reach densities higher than the ones obtained in the linear $\sigma$-model.
The \textit{Wigner-Seitz approximation}~\cite{Wigner:1933zz}, coming from solid state physics, has been used as a 
useful tool to mimic a dense system for both non-topological soliton models~\cite{Birse:1987pb, Reinhardt:1985nq, U.Weber1998}, 
chiral soliton models~\cite{Glendenning:1986fy, Hahn:1987xr, U.Weber1998, P.Amore2000} and 
also the Skyrme model~\cite{Wuest:1987rc, Amore:1998fp}.  The work in the Skyrme model reported above 
when compared with that of the Wigner-Seitz approximation shows the limitations of the latter, e.g., long range correlations, 
implementation of isospin rotations, etc... We would like ultimately to implement, using the CDM, the dense matter description 
in a real volume  to understand the role played by the quarks in the phase transitions. As we learned from the Skyrme investigation 
the first step is to understand in detail the long distance skyrmion-skyrmion interaction. Therefore, we present in here a detailed  study of the 
B=2 system in the CDM, as a first step in the attempt to go beyond the Wigner-Seitz approximation.

When the two solitons are  far apart the use of the \textit{product ansatz} leads to a good description of the interaction between  
two solitons which depends on the relative orientation of the hedgehog quills and in the case of  the Skyrme model  leading to a 
suitable description of the required interaction to describe dense matter \cite{Park:2003sd,Park:2008zg}.The main difference of the present approach 
with respect to the latter is the inclusion  of quark degrees of freedom coupled to non-topological solutions of the meson 
fields ~\cite{Blattel:1987ta}. The quarks in the product ansatz provide a strong repulsion at short distances due to the Pauli 
Principle, i.e. when the two baryons approach each other with hedgehog quantum numbers there are only three quarks states and therefore the six quarks do not fit into them leading to Pauli blocking.
Thus at short distances we look for another ansatz. The idea is to find a bound state of six quarks where the quarks are excited into 
higher lying levels. We use this state, allowing  the size to vary, as a variational ansatz for the $B=2$ system at short 
distances. We match the two solutions at intermediate distances  obtaining in a specific channel an approximately 8 MeV bound state. 

A more realistic  analysis of the $B=2$ system in a chiral quark-soliton model has already been performed  by Sawado and collaborators~\cite{Sawado:1998gk, Sawado:2000tf}. In these papers the authors provide a numerically solution for the $B=2$ system in the chiral quark soliton model introduced by Diakonov and others~\cite{Diakonov:1987ty, Wakamatsu:1990ud, Reinhardt:1988fz} and they show that the axially symmetric configuration of meson fields leads to the minimum energy configuration. At this early stage of our inspection of the phase transitions in skyrmionic matter, the numerical analysis of Sawado is difficult to implement.  

The structure of the paper is as follows. In Sec.~\ref{model} we
describe model used. In Sec.~\ref{b2} we present the calculations for the $B=2$ system in the long/intermediate distance regime.
Sec.~\ref{shortdist} is dedicated to study the short distance calculation with the six bag ansatz 
and present the results.
 Finally, in Sec.~\ref{conclusions}
we summarize our findings  and present future outlooks.

\section{The Chiral Dilaton Model}\label{model}

The results presented in this section have already been discussed in detail by ref.~\cite{Drago:2011hq}.
Here the authors studied the $B=1$ solutions of the CDM and besides providing a finite density analysis through the Wigner-Seitz approximation, they also examined the solitonic solutions of the model in vacuum. We present their development to introduce the notation and provide all the required background for the description of the $B=2$ systems.

The model in ref.~\cite{Drago:2011hq} is defined by the following lagrangian density~\cite{E.K.Heide1994,G.W.Carter1998,G.W.Carter1997,G.W.Carter1996,Bonanno:2008tt}, i.e.
a simplified chiral Lagrangian which contains only chiral fields coupled to quarks,
\begin{align}\label{lagrCDMchiral}
\mathcal{L}  = \bar{\psi}  [i \gamma ^\mu\partial _{\mu}-
g_\pi(\sigma + i \boldsymbol{\pi}\cdot \boldsymbol{\tau}\gamma _{5})]\psi 
  +\dfrac{1}{2}(
\partial_{\mu}\sigma \partial^{\mu}\sigma + \partial_{\mu}\boldsymbol{\pi}\cdot \partial^{\mu}\boldsymbol{\pi})
-V(\sigma,\mathbf{\pi}).
\end{align}
The potential is given by:
\begin{align} \label{potfro}
& V(\sigma,\bb{\pi}) =  \lambda_1 ^2 (\sigma^2 +\boldsymbol{\pi}^2)-\lambda_2 ^2\,\ln(\sigma^2 + 
\boldsymbol{\pi}^2)-\dfrac{\epsilon_1}{\sigma_0} \sigma
\end{align}
where:
\begin{align}
& \lambda_1 ^2=\dfrac{1}{2}\dfrac{B \delta \phi_0 ^4+\epsilon_1}{\sigma_0 ^2}=\dfrac{1}{4}(m_\sigma ^2+m_\pi ^2)\\
& \lambda_2 ^2=\dfrac{1}{2}B \delta \phi_0 ^4= \dfrac{\sigma_0 ^2}{4}(m_\sigma ^2-m_\pi ^2).
%& \epsilon_1 =m_\pi ^2 \sigma_0 ^2.
\end{align}
Here $\sigma$ is the scalar-isoscalar field, $\bb{\pi}$ is the pseudoscalar-isotriplet meson field, $\phi$ is
the dilaton field which, in the present calculation, is kept frozen at its vacuum value $\phi _0$ and $\psi$ describes the isodoublet quark fields.\\
The constants B and $\phi_0$ in the potential are fixed by choosing a value
for the mass of the glueball and for the vacuum energy,
while $\delta = 4/33$ is given by the QCD beta function and it
represents the relative weight of the fermionic and
gluonic degrees of freedom. 
More details on the construction of this logarithmic potential can be found in the works of Schechter~\cite{J.Schechter1980} and of Migdal and Shifman~\cite{Migdal:1982jp}.

The decision to keep the dilaton field frozen is based on the results
obtained in~\cite{Bonanno:2008tt, G.W.Carter1998}, where it has been
shown that at low temperatures the dilaton remains close to its vacuum
value even at large densities. 

The vacuum state is chosen, as usually, at $\sigma_0=f_\pi$ and $\bb{\pi}=0$.
This Lagrangian density, besides being invariant under  chiral symmetry, is also invariant under another fundamental
symmetry in QCD, scale invariance, which is also spontaneously broken,  and is  implemented above following Refs.~\cite{J.Schechter1980,Migdal:1982jp,E.K.Heide1992}.
In the following calculations we use the parameters which have served to fit the B=1 systems: $g_\pi=5$, $m_\sigma=550$ MeV, $m_\pi=139$ MeV and $f_\pi=93$ MeV.\\

The self-consistent $B=1$ solution in the CDM obtained by adopting the hedgehog ansatz, plays the role of the skyrmion in Skyrme's model, i.e., the true baryonic states are obtained from it by quantization.
For the Lagrangian in eq.~(\ref{lagrCDMchiral}) the hedgehog ansatz is defined by
\begin{align}\label{hedgenansatz}
& \sigma_{B=1} (\vec{r}) = \sigma_h (r) \;\; \; ,     \;\; \;  \vec{\pi}_{B=1} (\vec{r})= h(r) \hat{r}\nonumber \,\, ,\\
& \psi_{B=1} (\bb{r}) = \frac{1}{\sqrt{4 \pi}}
\left(\begin{array}{c} 
u(r) \\ 
i \bb{\sigma} \cdot \hat{\bb{r}} v(r)
\end{array} 
\right)
 \frac{1}{\sqrt{2}} (| u \downarrow \rangle - | d \uparrow \rangle).
\end{align}
 The field equations for the Dirac components become,
\begin{align}
& \dfrac{du}{dr}=g_\pi hu +\left (-\epsilon -g_\pi \sigma_h \right)v\,\,\, ,\\
& \dfrac{dv}{dr} =-\dfrac{2}{r} v -g_\pi h v+\left(\epsilon -g_\pi \sigma_h \right)u\,\,\, , 
\end{align}
where $\epsilon$ is the eigenvalue for the quark spinor $\psi_{B=1}$.\\
For the chiral fields, the equations become,
\begin{align}
& \dfrac{d^2\sigma_h}{dr^2}= -\dfrac{2}{r}\dfrac{d\sigma_h}{dr}+\dfrac{3g_\pi}{4\pi}(u^2 -v^2)+\dfrac{\partial V}{\partial \sigma_h}\\
& \dfrac{d^2 h}{dr^2}=-\dfrac{2}{r}\dfrac{dh}{dr}+\dfrac{2}{r^2}h
+\dfrac{3g_\pi}{2 \pi}uv +\dfrac{\partial V}{\partial h}.
\end{align} 

\begin{figure}[h!]
\begin{center}
 \includegraphics*[width=0.45\textwidth]{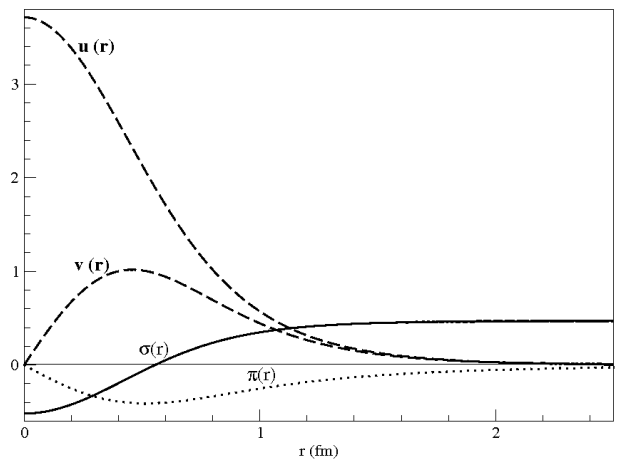}
\caption{We plot the fields in the hedgehog configuration for the given parameters (results taken from~\cite{Drago:2011hq}). Dirac fields $u(r)$ and $v(r)$ are in $fm^{-3/2}$, while the sigma and the pi mesons are expressed in $fm^{-1}$. The $\sigma$ field defines a bag like structure and the quarks fields are relevant only inside.The
solitonic behaviour of the pion field for large distances, despite the fact that the theory has trivial topology, is apparent.
}
\label{hedgehogB1}      
\end{center}
\end{figure}

The solution has been obtained by imposing the following boundary conditions at $r=0$,
\begin{eqnarray}
& u'(0)=v(0)=0\nonumber ,\\
&\sigma_h '(0)=h (0)=0,
\end{eqnarray}
and at infinity,
\begin{align}
& \sigma_h (\infty)=f_\pi ,\,  h (\infty)=0,\nonumber\\
& \dfrac{v(\infty)}{u(\infty)}=\sqrt{\dfrac{-g f_\pi+\epsilon}{-g f_\pi-\epsilon}}
\end{align}
where $\epsilon$ is the quark eigenvalue. We show in Fig. \ref{hedgehogB1}  the fields for the hedgehog configuration for the parameters mentioned before. The $\sigma$ field acts as an approximate order parameter defining a bag like structure. The quark fields are only sizeable inside this bag  of radius $0.5$ fm.
The pion field has a solitonic behaviour from $0.5$ fm on which resembles the structure of the skyrmion at large distances.   

The energy of this ``soliton" like solution  is given by,
\begin{align}\label{enB1}
E_{B=1}= 4\pi \int r^2 dr (E_{int}(r)+E_{kin,Q}(r)+E_{\sigma}(r)+E_\pi (r)+ E_{pot}(r))
\end{align}
where the quark-mesons interaction and the quark kinetic energies are determined by,
\begin{eqnarray}
& E_{int}(r)= \dfrac{3}{4\pi}\left[ g_\pi \sigma_h (u^2-v^2)+2g_\pi h uv \right]\\
& E_{kin,Q}(r)= \dfrac{3}{4\pi} \left(u \dfrac{dv}{dr}-v \dfrac{du}{dr}+\dfrac{2}{r}uv\right)
\end{eqnarray}
and the energy density of the meson fields and of the potential read,
\begin{eqnarray}
E_{\sigma}(r)&=& \dfrac{1}{2}\left( -\dfrac{d\sigma_h}{dr}\right)^2\\
E_\pi (r)&=&\left[\left(\dfrac{dh}{dr}\right)^2+ \dfrac{2}{r^2}h^2\right] \\
E_{pot}(r)&=&V(\sigma_h , h)
\end{eqnarray}

In Table~\ref{contrib} we show all the contributions to the total energy of the soliton, coming from mesons and quarks, for the set of parameters: $m_\sigma=550$ MeV, $m_\pi=139$ MeV and $g=5$.

\begingroup
\begin{table}[h]
\caption{Contributions to the soliton total energy at mean-field 
level. All quantities are in MeV (results taken from~\cite{Drago:2011hq}).}\label{contrib}
\vskip 0.2cm
\centering
\begin{tabular}{|c c c |c|}
\hline 
 & & & \\
Quantity &  &  & Our Model  \\[7pt]
 \hline
 & & & \\
Quark eigenvalue &  &  & $83.1$ \\[6pt]
Quark kinetic energy &  &  & $1138.0$\\ [6pt] 
$E_\sigma$ (mass+kin.)&  &  & $334.5$\\[6pt]
$E_\pi$ (mass+kin.)&   &  & $486.0$ \\ [6pt]
Potential energy $\sigma -\pi$& & & $105.7$\\ [6pt]
$E_{q\sigma}$&   &   & $-101.4$ \\ [6pt]
$E_{q\pi}$ &   &  & $-787.0$ \\ [6pt]
Total energy  &   &  & $1175.6$ \\ [10pt]  
 \hline
\end{tabular} 
\end{table}
\endgroup

We note that, as it happens with the skyrmion, for physical parameters the mass of this state is between the  nucleon and the Delta states. Quantization will lead to the approximate right masses.

\section{The soliton-soliton interaction}\label{b2}

\begin{figure}[b]
\begin{center}
 \includegraphics*[width=0.3\textwidth]{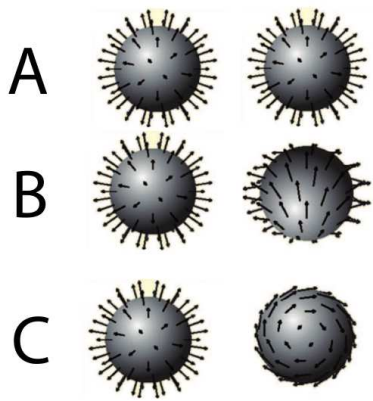}
\caption{The three configurations A, B, and C corresponding to the different orientation in the isospin space.}
\label{configurations}      
\end{center}
\end{figure}

The first step in order to understand dense matter in this model is to study the soliton-soliton interaction as a function of the  distance  between solitons. To do so we start by constructing a product ansatz solution valid for large distances.  A similar calculation has been carried out in the Skyrme model~\cite{Park:2003sd} and we will compare both to see the effect of the  fermionic degrees of freedom in this description.

The hedgehog configuration of the fields in eqs.(\ref{hedgenansatz}) also contains isospin degrees of freedom and in principle different configurations of solitons in the isospin space can lead to lower energy states in the lattice.
We focus on the study of the interaction of the two solitons by changing the relative orientation of the hedgehog quills in the regime of long distances, through the use of the \textit{product ansatz}.
This approach permits to describe the soliton-soliton interaction in the range of long/intermediate distances, namely  while the two $B=1$ solitons do not  strongly overlap.

Assume we have two solitons, each a $B=1$ hedgehog solution, whose centers are at $\bb{r}_1$ and $\bb{r}_2$. 
For simplicity we place the centers of the two hedgehogs symmetrically along the $\hat{z}$ axis at a distance $d$, hence the explicit expressions for the centers become:
\begin{align}
&\bb{r}_1=(0,0,-\dfrac{d}{2})\\
&\bb{r}_2=(0,0,\dfrac{d}{2}).
\end{align}
We define the displaced meson fields centered in each soliton as following:
\begin{align}\label{displfields}
& \sigma(\bb{r}-\bb{r}_1)= \sigma_1 \,\, , \sigma(\bb{r}-\bb{r}_2)= \sigma_2 \,\, , \nonumber\\
& \bb{\pi}(\bb{r}-\bb{r}_1)= \bb{\pi}_1 \,\, , \bb{\pi}(\bb{r}-\bb{r}_2)= \bb{\pi}_2 \,\, .
\end{align}
In the product ansatz scheme the new meson fields configurations read:
\begin{align}\label{chirfieldsB2}
& \left(\dfrac{\sigma_{B=2}(\bb{r})+i\bb{\tau}\cdot\bb{\pi}_{B=2}(\bb{r})}{f_\pi}\right)\nonumber\\
&  =\left(\dfrac{\sigma_1 +i\bb{\tau}\cdot\bb{\pi}_1}{f_\pi}\right) A \left(\dfrac{\sigma_2 +i\bb{\tau}\cdot\bb{\pi}_2}{f_\pi}\right) A^\dagger
\end{align}
where $f_\pi$ is introduced in order to have the proper dimensions and to recover the correct asymptotic behaviour.
The operator $A$ is a $SU(2)$ matrix inserted to take
into account the relative orientation in the isospin space of one of the solitons with respect to the other. 
More explicitly the meson fields for the $B=2$ system are given by:
\begin{align}\label{ExplchirfieldsB2}
\sigma_{B=2}(\bb{r}_1,\bb{r}_2)= & \dfrac{1}{f_\pi}\left[\sigma_1 \, \sigma_2 +\bb{\pi}_1\cdot \bb{\pi}_2\right]\nonumber\\
\bb{\pi}_{B=2}(\bb{r}_1,\bb{r}_2)= & \dfrac{1}{f_\pi}\left [\sigma_1 \bb{\pi}_2+\sigma_2 \bb{\pi}_1+\bb{\pi}_1\times \bb{\pi}_2\right ]
\end{align}
In the Skyrme model, taking the product of two $B=1$ soliton solutions, is one of the most convenient ways to 
obtain the $B=2$ intersoliton dynamics \cite{Lee:2003aq}. 
The relevant difference between the CDM and Skyrme model is the presence of quark degrees of freedom, that also needs to be taken into account in the new  fields configuration.
At zeroth order, since the meson background configuration shows a reflection symmetry except for an isospin rotation, the quark fields can be  expressed as a linear combination given by:
\begin{equation}\label{quarkB2}
\psi_{\pm} (\bb{r})=\dfrac{1}{\sqrt{2}}(\psi_L (\bb{r})\pm \psi_R (\bb{r}))\,\, ,
\end{equation}
where the left and right spinors read:
\begin{align}
& \psi_L(\bb{r})=\psi_1 (\bb{r}-\bb{r}_1)\,\, ,\nonumber\\
& \psi_R(\bb{r})=A\psi_2 (\bb{r}-\bb{r}_2).
\end{align}

In order to study the soliton-soliton interaction as a function of the isospin, we decide to keep one soliton fixed and rotate the other, namely the one centered in $\bb{r}_2$.
We will consider three different configurations, shown in Fig.~\ref{configurations}:
\begin{enumerate}
\item Configuration A: $A=\mathbb{I}$, i.e. the two solitons are unrotated (panel A in Fig.~\ref{configurations}).
\item Configuration B: $A=e^{i\frac{\tau_z}{2}\pi}=i\tau_z$, which corresponds to rotate the second soliton by an angle $\pi$ about the axis parallel to the line joining the two centers (panel B in Fig.~\ref{configurations}).
\item Configuration C:$A=e^{i\frac{\tau_x}{2}\pi}=i\tau_x$, leading to a rotation of $180$ degrees around the axis perpendicular to the line joining the two solitons, in our case the $x$ axis (panel C in Fig.~\ref{configurations}).
\end{enumerate}

\begin{figure}[htb]
\begin{center}
 \includegraphics*[width=0.63\textwidth]{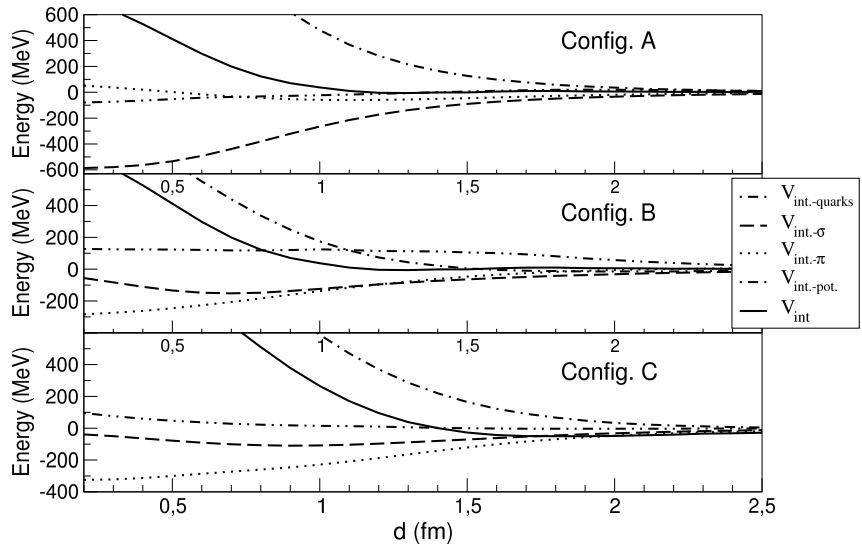}
\caption{Contributions to the soliton-soliton interaction energy as a function of the distance $d$ for the configurations $A$, $B$ and $C$. The solid black curve represents the total interaction energy defined in eq.~(\ref{intenergy}).}
\label{contribene}      
\end{center}
\end{figure}

\begin{figure}[t]
\begin{center}
 \includegraphics*[width=0.5\textwidth]{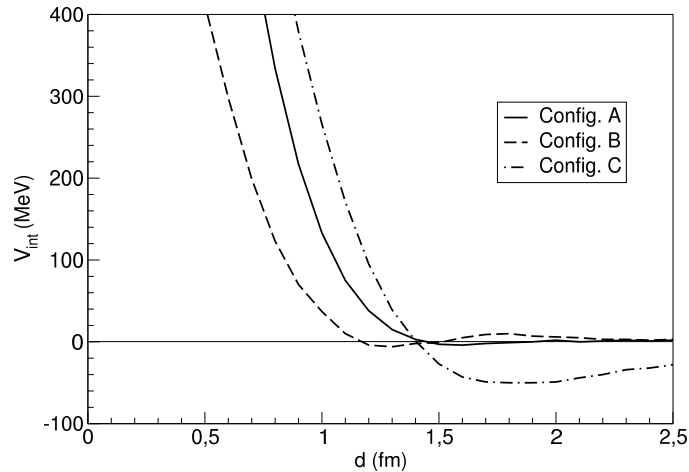}
\caption{Soliton-soliton interaction energy as a function of the distance $d$ for the three configurations A, B and C.}
\label{potential}      
\end{center}
\end{figure}

In the three configurations A, B and C just one pion is rotated, so we can interpret this transformation as a changing in the isospin orientation of the pion. 
The quantum numbers of the pion are $S=0$, $I=1$ and intrinsic parity $P=-1$, leading to a grand-spin $G^P=1^-$ which admits three different projections $G_3=-1,0,1$. Hence by rotating the pion field we obtain a linear combination of $G_3$ projections.\\
The total grand-spin for the $B=2$ system is given by the sum  of the grand-spin $G_1$ for the soliton placed in $r_1$ and of $G_2$ for the one centred in $r_2$.
Moreover we have to take into account the total parity $P_{tot}$ of the system. For the quark state, since all quarks lie in the $0^+$ state, they possess positive parity. The pion couples to the spinor through the $\gamma_5$ matrix, which under parity changes sign, thus the globally parity carried by the pion coupling will again be positive. This leads to a total parity $P_{tot}$ equal to $+1$.
The total grand-spin will follow from:
\begin{align}
& G_1^{P_1}=G_{1,quarks} ^+ +G_{1,pion}^+=1^+&\nonumber\\
& & \Longrightarrow G_{tot}^{P_{tot}}=0^+,1^+,2^+\\
& G_2^{P_2}=G_{2,quarks} ^++G_{2,pion}^+=1^+&\nonumber
\end{align}
In particular, from eqs.~(\ref{ExplchirfieldsB2}), it can be seen that the product ansatz represents a possible way to write the pion component of the $B=2$ solution as a vector which is associated to the quantum number $G=1$.
The difference between the three configurations lies in the different spatial-isospin arrangement of the fields. For instance the only configuration that keeps unaltered the hedgehog structure when the two solitons are on top of each other is the $A$ configuration (for which $G_{3,tot}=0$). The remaining configurations $B$ and $C$, due to the rotations, will generate an orientation of the chiral fields which will give a different correspondence between spatial and isospin degrees of freedom.

\begin{figure}[b]
\begin{center}
 \includegraphics*[width=0.45\textwidth, angle=-90]{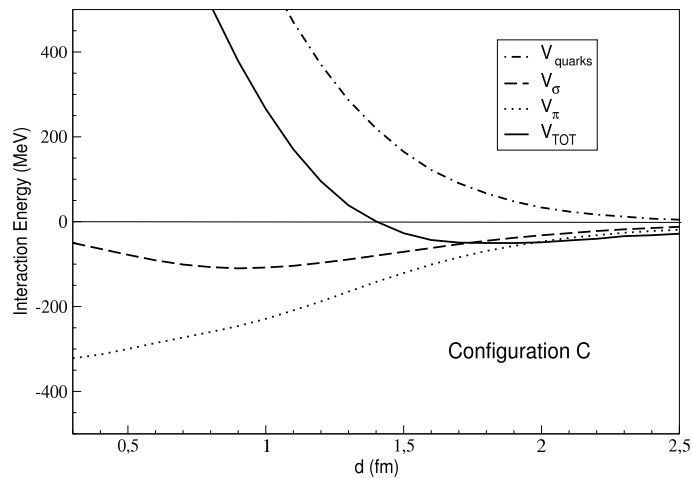}
\caption{Soliton-soliton interaction energy as a function of the distance $d$ for the configuration C. The dotted curve represents the contribution  to the interaction energy of the $\pi$ field, the dashed curve corresponds to the contribution of the $\sigma$ field and the dashed-dotted curve to that of the quarks fields. The full curve represents the total interaction energy.}
\label{quarkmeson}      
\end{center}
\end{figure}

For each isospin configuration (A,B,C) we calculate the energy of the $B=2$ soliton using the approximated field expressions given in~(\ref{ExplchirfieldsB2}) and  evaluating the expectation value of the Hamiltonian on the new $B=2$ state where the valence quarks fill the levels provided by eq.~(\ref{quarkB2}).
The energy hence reads:
\begin{equation}\label{sumcon}
E_{B=2}=E_{\sigma, B=2}+E_{\pi, B=2}+E_{pot, B=2}+E_{Q, B=2}.
\end{equation}

The soliton-soliton interaction energy as a function of the intersoliton separation $d$ is defined as:
\begin{equation}\label{intenergy}
V_{int}(d)=E_{B=2}-2E_{B=1}
\end{equation}
where the energy for the $B=1$ system is given by eq.(~\ref{enB1}).\\

In Fig.~\ref{contribene} we show for each configuration the different contributions to the total energy, defined in eq.~(\ref{sumcon}), subtracted by twice the corresponding $B=1$ energy. In this way we are able to evaluate the changes in the interaction energies from the $B=1$ case to the $B=2$ system in the product ansatz approach.\\
 It can be seen that the differences between the three isospin configurations arise only at short distances.
The meson interaction energies for the $\sigma$ and the $\pi$ field always show an attractive behaviour; in particular the configuration $A$, where the two solitons end up sitting on the top of each other, is the most attractive for the sigma, 
while for the pion the maximum attraction occurs for the configuration $C$.
On the other hand the quark contribution is mostly repulsive and quite large in all the isospin arrangements.
The potential energy evaluated in the $B=2$ system does not  change significantly from the one-soliton calculation.\\

In Fig.~\ref{potential} we present the total interaction energy for the three configurations. 
There is no significant difference in shape between the three cases but the three curves show very different behaviours as $d$ decreases.
First of all we notice that at long/intermediate distances the $B=1$ case is recovered, as expected since the two solitons are well localized and far apart.
As the separation reduces the quark Pauli repulsion starts acting and at $d\lesssim 1$ fm all the configurations show a strong repulsive trend. The A configuration is basically repulsive, showing only  a small attraction of a few MeV  around 
$d= 1.5$ fm. The B configuration is also fundamentally repulsive except for a small attraction at smaller distances, $d=1.2 $ fm. Finally the C configuration is strongly attractive above $d= 1.4$ fm  allowing for a strongly deformed bound state.

The contribution of the quark fields plays a crucial role at short separation. The large repulsion is due to the fact that we are forcing the six quarks to stay in a Grand-Spin $G^P =0^+$ and does not allow to get a clear insight in the short range regime. This is shown explicitly in In Fig.~\ref{quarkmeson} for configuration C. The same quark repulsion mechanism is operative for configurations A and B. The C configuration, thanks to the $\pi$ and $\sigma$ contributions, is the most attractive.

\begin{figure}[h!]
\begin{center}
 \includegraphics*[width=0.44\textwidth]{potential.eps} 
 \includegraphics*[width=0.44\textwidth]{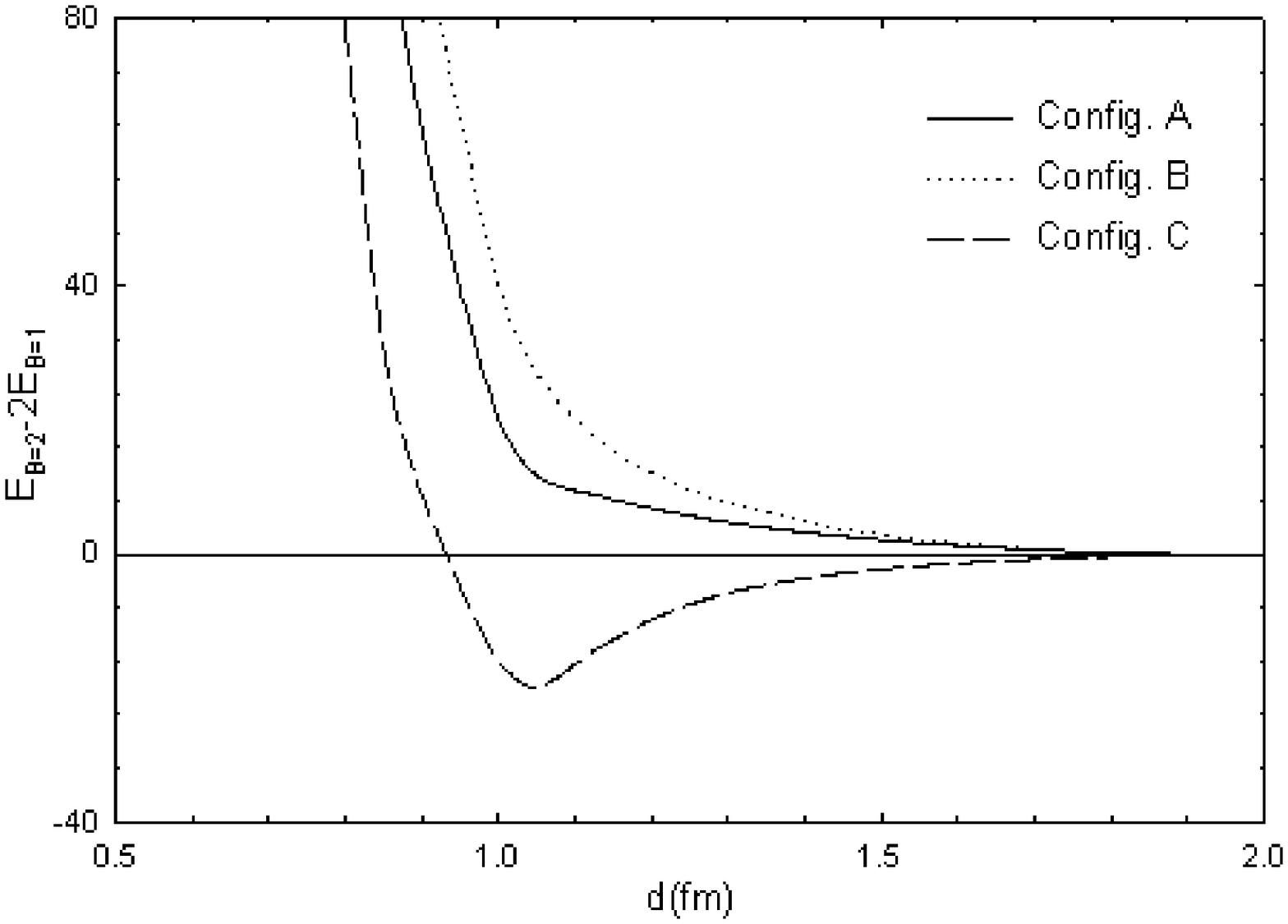}
\caption{We compare the CDM soliton-soliton interaction (left) with the skyrmion-skyrmion interaction~\cite{Park:2003sd} (right) energy  for the configurations A,B,C in a Skyrme model with vector $\omega$ and $\vec{\rho}$ mesons. The repulsion in this latter case is produced by the vector mesons. The strong attraction in channel C is provided by a combination of the skyrmion and the $\omega$ meson.}
\label{skyrmeVM}      
\end{center}
\end{figure}

In Fig~\ref{skyrmeVM} we compare our result with that of the Skyrme model with vector mesons \cite{Park:2003sd}. From our experience there in the Skyrme model, we note that the mechanism for the phase transitions strongly depend on the attractive piece of the interaction. The short range repulsion affects only the details of the process. Therefore, we expect that the $C$ configuration is the one that will lead to the phase transition as in the Skyrme case~\cite{Lee:2003eg,Park:2008zg,Park:2009bb}. Thus, in order to describe the behaviour of nuclear matter we should formulate a scheme based on the $C$ interaction configuration.

But before we do so we would like to understand how one can describe the short distance behaviour of the two body interaction, useful for the description of finite nuclei, in the hedgehog approximation to the CDM.

\section{The short distance behaviour}\label{shortdist}

The interaction at short distances cannot be described by the product ansatz since one has to take into account  the Pauli effect associated to Fermi statistics. In order to do so we need to  include the possibility that quarks move into excited states, since as the Pauli blocking effect becomes large, quarks will be favoured to populate excited states.
Thus, as the distance between the two solitons reduces,  the system would preferably go into the lowest energy configuration in which the six quarks fill the appropriate quark levels
satisfying the Pauli principle. We  proceed to show an analytic calculation in the hedgehog approximation of the CDM.

\begin{figure}[t]
\begin{center}
 \includegraphics*[width=0.44\textwidth]{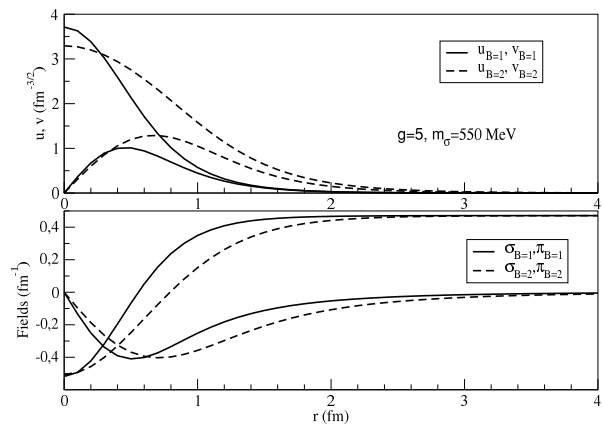} 
\caption{Upper panel: Dirac component profiles for the $B=1$ solution (solid line) and the six quarks bag (dashed line). Lower panel: sigma and pion fields for the one soliton case (solid line) and the six quarks bag (dashed line).}
\label{B2fields}      
\end{center}
\end{figure}

At zero separation the $B=2$ bound state  system is given by the self-consistent solution of the Lagrangian in eq.~(\ref{lagrCDMchiral}) with six quarks. The exact solution of this six quarks bag provides the energy of the $B=2$ system when $d=0$.
Following ~\cite{Blattel:1987ta}, we solved self-consistently the field equations for the model with six quarks by assuming the hedgehog ansatz for the fields. 
The hedgehog structure requires a symmetric occupation of the third components of the Grand-Spin $G$; from this constraint the source terms for the $\sigma$ and  $\pi$ fields are invariant and the field equations will differ from the three quarks case just in some multiplying constants. The quarks and mesons field profiles do not change significantly from the $B=1$ case as can be seen in Fig.~\ref{B2fields}.
It should be noticed that the $B=2$ configuration presents an enhanced long range behaviour in the quark and meson field profiles that leads to a wider bag-structure for the sigma field. Nevertheless, since the scalar density of quarks keeps the same shape as in the $B=1$ case, the sigma field does not undergo a deep modification.
Moreover this behaviour is in agreement with the results shown in ~\cite{Blattel:1987ta}.
From these considerations, we can anticipate that, in order to preserve the hedgehog configuration, the matching between the short and the long range interactions will only involve the $A$ configuration of the product ansatz. The $B$ and $C$ configurations will require more complex $B=2$ bound states which are not of hedgehog type. The $C$ configuration is strongly bound by the intermediate  range behaviour of the $\sigma$ and therefore corresponds most probably, when quantized, to the deuteron.

Let us proceed with the $A$ configuration. The ground state of the hedgehog solution is given by:
\begin{equation}
\vert G^P=0^+\rangle = \vert G=G_3=0, j=\frac{1}{2}, l=0,k=-1\rangle \nonumber
\end{equation} 
where $j,l, k$ are respectively the total angular momentum, the orbital angular momentum and the Dirac quantum number. Here we can put three quarks since we include the colour quantum number.
The next level is the state $\vert G^P=0^-\rangle =\vert G=G_3=0,j=\frac{1}{2},l=1,k=1\rangle$ and again here we can put the other three quarks.
The next state in energy available is the $\vert G^P=1^+\rangle$ which can accept up to nine quarks since $G_3=-1,0,1$.
So we  have two possibilities to place six quarks:
\begin{itemize}
\item[ i)] $0^+,0^-$ with total grand-spin $G_{tot}^{P_{tot}}=0^-$;
\item[ ii)] $0^+,1^+$ with total grand-spin $G_{tot}^{P_{tot}}=1^+$.
\end{itemize}

As already shown in~\cite{Blattel:1987ta} the set that provides the lowest energy state is ii), where we place three quarks in $1^+$ and the other three quarks in $0^+$ \footnote{As already anticipated, to maintain the hedgehog structure we place one quark in each one of the states $G_3=1,0,-1$.} .

The total energy for the six quarks bag reads:
\begin{align}
E&= N_{G=0}  \epsilon_0+N_{G=1} \epsilon_1 + E_\sigma +E_\pi +E_{pot}\nonumber\\
&= 3(\epsilon_0+\epsilon_1)+ E_\sigma+E_\pi +E_{pot}
\end{align}
where $\epsilon_0, \,\epsilon_1$ are respectively the quark eigenvalues for the $0^+$ and $1^+$ states.
In Table~\ref{energysix}  we show the results for the values $m_\sigma=550$ MeV and $g=5$.
we also present the interaction energy $V_{int}=E_{6quarks}-2 E_{B=1}$

\begingroup
\begin{table}[h]
\caption{Contributions to the total energy. All quantities are in MeV.}\label{energysix}
\vskip 0.2cm
\centering
\begin{tabular}{|c c c |c|}
\hline 
 & & & \\
Quantity &  &  & Our Model \\[7pt]
 \hline
 & & & \\
$\epsilon_0$ &  &  & $-30.6$ \\[6pt]
$\epsilon_1$ &  &  & $344.8$ \\[6pt]
$E_\sigma$ (mass+kin.)&  &  & $470.6$\\[6pt]
$E_\pi$ (mass+kin.)&   &  & $643.8$ \\ [6pt]
Potential energy $\sigma -\pi$& & & $287$\\ [6pt]
Total energy  &   &  & $2344$ \\ [6pt]  
Total energy $B=1$ &   &  & $1175.6$ \\ [6pt]  
 \hline
 & & & \\
$V_{int}$ &    &    & $-7.2$ \\[10pt] 
\hline
\end{tabular} 
\end{table}
\endgroup

The next step is to connect the result at $d=0$ with the results coming from the product ansatz at large distances. 
We showed that the $B=2$ system, written through the product ansatz, admits $G_{tot}^{P_{tot}}=1^+$; in addition the solution given by the $A$ configuration is the only one that keeps a hedgehog structure. Hence we expect that as $d$ decreases the system will move from the configuration $A$ in the product ansatz to the state provided by the six quarks bag.

In order to obtain the energy of the six quarks bag as a function of the intersoliton separation, we adopt a variational approach, namely we use the exact solutions for the six quarks problem as initial ans\"{a}tze and evaluate the total energy at each value of $d$.
The trial wave functions (following~\cite{Birse1985}) for the meson fields read:
\begin{align}
& \sigma (r, R_c)=-f_\pi \cos \left[\pi \tanh \left (\dfrac{\log 3}{2 R_c}r \right )\right]\,\, ,
& \pi (r,R_c)=-f_\pi \sin \left[\pi \tanh \left (\dfrac{\log 3}{2 R_c}r \right )\right]\,\, ,
\end{align}
where the variational parameter $R_c$ is related to the range of action of the attractive potential of the chiral fields.
The appropriately normalized Dirac components are given by the following expressions:
\begin{align}
& u(r,R_D)= \dfrac{1}{\sqrt{N(R_D)}} u_0 \exp \left[ -\dfrac{r^2}{R_D ^2}\right ]\,\, ,
& v(r,R_D)= \dfrac{1}{\sqrt{N(R_D)}} v_0 r \exp \left[ -\dfrac{r^2}{R_D ^2}\right ]\,\, .
\end{align}
The parameter $R_D$ is introduced in order to obtain a relation between the radius of the bag, which involves the presence of the Dirac components defined above, and the inter-soliton distance $d$:
%Here the parameter $R_D$ is strictly connected to the size of the bag, and hence to the distance $d$, through the relation:
\begin{equation}\label{radbag}
d=(<r_{bag} ^2>)^{1/2}(R_D)=\left (\int  r^4 (u^2 (r,R_D)+v^2 (r,R_D))dr \right) ^{1/2}.
\end{equation}

The total energy of the bag depends on the variational parameter $R_c$, $R_D$ and on the quark-chiral fields coupling $g$ and is given by:
\begin{equation}\label{en1}
E(R_c,R_D,g)= E_{int}(R_c,R_D,g)+E_{kin.quarks} (R_D)+E_\sigma (R_c) +E_\pi (R_c) +E_{pot} (R_c)\nonumber
\end{equation}
where $E_{int}(R_c,R_D,g)$ represents the quark-meson interaction energy.
As already mentioned previously, the interaction energy for the six quarks bag as a function of the distance is defined as:
\begin{align}
V_{int,6}(d)=E(R_c,d(R_D),g)-2 E_{B=1}.
\end{align} 
The variational method adopted to calculate $V_{int,6}$ consists in fixing $g=5$ and varying $R_c$ to minimize the energy functional in eq.(~\ref{en1}) at each value of $R_D$.

\begin{figure}[h]
\centering
\includegraphics*[width=0.55\textwidth]{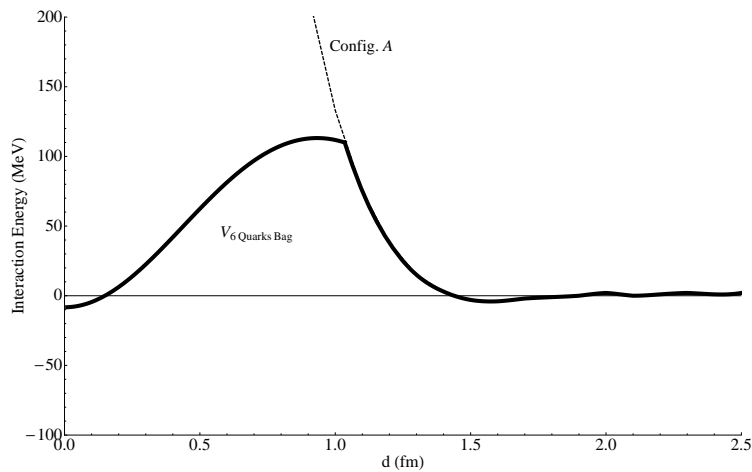}\caption{Interaction energies as function of the intersoliton distance; here we show the $A$ configuration obtained with the product ansatz approach (dashed lines) and the results from the six quarks bag for $R_c=1.04$ (black solid line)}\label{crossingall}
\end{figure}

In Fig.~\ref{crossingall} we present the interaction energy as a function of the distance $d$ between the two solitons.
The behaviour of the $B=2$ system moving from large to small separations, is as follows:
\begin{itemize}
\item at distances $d\gtrsim 1.5$ fm, the $A$ configuration, as expected, provides a solution which is basically the sum of the two $B=1$ solitons; 
at large separation the behaviour of the soliton interaction is similar to that obtained in the Skyrme model since when two solitons are far apart, the interaction between them is 
mostly mediated by the chiral fields;

\item as $d$ is reduced, the quark repulsion of the product ansatz  makes the interaction repulsive until it matches that of the six quark bag, thereafter the system undergoes a transition from $A$ to the six quarks bag configuration at roughly $d_{A6} \approx 1$ fm and the system becomes bound.
\end{itemize}

At intermediate distances, around $d \approx 1.5$ fm, the $A$ configuration is lightly attractive by a few MeV, but as $d$ gets smaller the quark repulsion increases and the interaction energy starts to rise until a crossing between the $A$ configuration and the solution of the six quarks bag occurs. This takes place for $d \sim 1$ fm.
For lower values of $d$ the systems goes into a stable configuration  of six quarks with $R_c \approx 1$ fm. From then on the $B=1$ hedgehogs are captured to form a quasi bound state leading at zero separation to the local minimum. Thus as the inter-soliton separation shrinks the system falls into a six quarks bag bound state where the inclusion of excited states for the quarks has produced an attractive configuration.
It should be noted that the optimal value of the variational parameter $R_c$, around $1$ fm, is compatible with the usual size of a soliton; moreover, since the transition to the six quarks bag  occurs at a value of $d$ compatible with $R_c$,  we notice that the overlapping of chiral fields does not affect  strongly the short range behaviour, which is completely determined by the quark contribution.  

\section{Conclusions} \label{conclusions}

We have used a Lagrangian with quark degrees of freedom based on chiral and scale 
invariance to study the soliton-soliton interaction with the aim set in nuclear matter studies.
We have described how this interaction energy behaves as a function of the distance $d$ between the two solitons.
When the two solitons are  far apart  we use a product ansatz solution. This method allows to  take into account the isospin degrees of freedom of the fields and permits an evaluation of the energy of the $B=2$ system for different isospin configurations.
In the long/intermediate distances regime, the system admits a strongly attractive configuration (C) corresponding to one soliton  rotated $\pi$ along an axis orthogonal to the line joining the two centers of the $B=1$ hedgehogs. Our experience in skyrmion matter teaches us that this is the solution which will govern the phase diagram in nuclear matter. It arises due  to a strong attractive  interaction associated with the $\sigma$ and $\pi$ fields. We expect that this solution will lead to the deuteron after quantization.The other two configurations are very slightly attractive to allow for bound states at long/intermediate hedgehog separations.

In the product ansatz calculation, by reducing the inter-soliton distance we are actually forcing six quarks to lie in the ground state with Grand-Spin $G^P=0^+$. Since this state can  accommodate only three quarks as the two solitons approach each other the three configurations develop a  repulsive core associated to the overlap of the $B=1$ solutions and due to the action of the Pauli principle for quarks.

The Skyrme analysis is based only on meson fields and therefore is blind for colour and for elementary fermionic degrees of freedom. Nevertheless, under the assumption that the confinement-deconfinement phase transition and  chiral symmetry restoration occur very close in the phase diagram, one can imagine beautiful scenarios for the dense phase of nuclear matter \cite{Park:2003sd,Park:2008xm,Park:2009bb}. The mechanism for the phase transition is dominated in these calculations by the long range tail of the pion (skyrmion) interaction, which is not the na\"{i}ve Yukawa potential since the pions (skyrmions) are solitons, therefore collective degrees of freedom. This long range tail is well described by the product ansatz \cite{Park:2003sd,Park:2008xm}.

The ultimate goal of  our investigation is to understand the behaviour of the fermionic degrees of freedom, the quarks, in the phase transition.  For this, the present project,  to understand the B=2 interaction, is fundamental. The idea was to have a two phase scenario, i.e. like the little bag for example, with a connection between the fermionic degrees of freedom and the mesonic degrees of freedom. However, such a scenario would be geometrically prohibitive. The CDM is a very nice instrument because it formulates a two phase picture in terms of a continuous field theoretic description without boundaries. The boundaries are in some sense smooth and generated by the sigma field. Note that the meson fields here have a very different nature as in Skyrme's analysis, since the baryon number is only lodged in the quarks.  Thus a half skyrmion phase will arise when the quark fields of the different baryons are delocalized and this can happen when the sigma field extends between several baryons as can be envisaged from  Fig. \ref{quarkmeson}.

We have found  a series of very interesting features in comparison with the skyrmion approach. The importance of the Pauli principle becomes relevant, so much that repulsion arises without the need of vector mesons. The behaviour of the interaction with isospin is quite similar to that of the skyrme approach. Again the C configuration is the most attractive one. No wonder, since the attraction is associated to the mesons. It will be precisely this channel where, after quantization, one expects to find the deuteron, since proton and neutron have different isospin projections. Moreover this is the channel that has to be used in nuclear matter studies.

In our investigation we have found two approximations which  lead to an almost exact solution for the B=2 problem  within the hedgehog ansatz. One approximation is the $A$ configuration which has $B=2$ hedgehog structure. The other we obtain by  constructing an  exact solution for the six quarks bag at zero separation solving self-consistently the field equations assuming  hedgehog ans\"atze for the fields. In this way we have found a stable six quarks bag bound state at $d=0$ with a binding  energy of $\approx -8$ MeV.
Starting from these exact field configuration, we have built an energy functional $V_{int,6}(d)$ and extended the results to finite values of  $d$ in order to reach the intermediate interaction range ( $d\approx 1$ fm) where the product ansatz description  still holds.
In order to maintain the hedgehog structure of the fields, used in the six quarks bag approach, the only configuration at large distances which  admits  hedgehog structure is the $A$ state, with no relative isospin orientation between the two solitons. 
We have presented the full description of the interaction energy as a function of the inter-soliton separation. The $B=2$ system, starting at large values of $d$ in configuration $A$, increases its energy as the distance shortens and the repulsion of the quarks becomes dominant. Around $d \sim 1$ fm  it becomes favourable for the quarks to jump into excited states leading to a six quarks bag which provides the most stable configuration and  finally the two solitons are captured forming a six quark hedgehog state.

We would like to stress the fact that the solution for the $B=2$ system presented here cannot represent a deuteron because we are not starting from a proton and a neutron, but the main components are two unprojected soliton states.  But, as we claim in this work, we expect that the binding configuration $C$, 
once projected, would lead to the deuteron. Nevertheless, this study, still at a hedgehog level, reveals the possible existence of bound 
states in a certain isospin configuration. The existence of these bound states in the different configurations, simply underlines the fact 
that the purely spherical configuration does not lead to the minimum energy, and the inclusion of isospin degrees of freedom allows 
to open new channels with lower energy configurations. 

The work presented here can be improved in several ways. The analysis of the $B=2$ system, still in this simplified version, can be extended to include configurations beyond the hedgehog ansatz, which could lead to lower energy state configurations, as already pointed out in previous works~\cite{Sawado:1998gk}.
The study of the $B=2$ system and more in general of a multi-soliton can be performed by building a three dimensional lattice with specific symmetries that leads to lower energy configurations~\cite{Brown:2009eh}. 
Moreover, one could include the dynamics of the dilaton  and perhaps also  vector mesons in the calculation in order to perform the study at finite density and temperature, as in Ref.~\cite{Park:2003sd,Park:2008xm,Park:2009bb}.

\begin{acknowledgements}
We would like to thank Alessandro Drago for useful discussions regarding hadron
structure and dense matter.\\
This work was supported by the INFN-MINECO agreement "Hadron structure in hot and dense medium".\\
Professor B.Y. Park was partially supported by the
WCU project of Korean Ministry of Education, Science
and Technology (R33-2008-000-10087-0). B.Y. Park, V. Vento and V. Mantovani Sarti have been also funded by the Ministerio de Economía y Competitividad 
and  EU FEDER under contract FPA2010-21750-C02-01, by Consolider Ingenio 2010
CPAN (CSD2007-00042) and by Generalitat Valenciana: Prometeo/2009/129. 
V. Mantovani Sarti was also partially supported by the IDAPP Program.

\end{acknowledgements}

\bibliography{biblio}
\bibliographystyle{h-physrev3}

\end{document}